\documentclass[aps,pre,twocolumn,longbibliography,unsortedaddress,floatfix]{revtex4-1}
\usepackage{amsmath,amssymb,amsfonts}
\usepackage{algorithmic}
\usepackage{graphicx} 
\usepackage{color} 
\usepackage{textcomp}

\usepackage{mathtools}
\usepackage{bm}
\usepackage{url}
\usepackage{ulem}
\newcommand{\bA}{\bm{A}}
\newcommand{\bC}{\bm{C}}

\newcommand{\bF}{\bm{F}}
\newcommand{\bJ}{\bm{J}}
\newcommand{\bU}{\bm{U}}
\newcommand{\bX}{\bm{X}}
\newcommand{\bZ}{\bm{Z}}
\newcommand{\bb}{\bm{b}}
\newcommand{\bp}{\bm{p}}
\newcommand{\bq}{\bm{q}}

\newcommand{\bxi}{\boldsymbol{\xi}}
\newcommand{\BSeta}{\boldsymbol{\eta}}
\newcommand{\bzeta}{\boldsymbol{\zeta}}

\newcommand{\bcalX}{\boldsymbol{\calX}}
\newcommand{\calX}{\mathcal{X}}

\newcommand{\ddt}[1]{\frac{d#1}{dt}}

\newcommand{\ddtheta}{\frac{d}{d\theta}}
\newcommand{\flow}[2]{\Psi_{#1,#2}}

\newcommand{\argmin}{\mathop{\rm argmin}\limits}

\newcommand{\tr}{\mathrm{tr}}

\def\BibTeX{{\rm B\kern-.05em{\sc i\kern-.025em b}\kern-.08em
    T\kern-.1667em\lower.7ex\hbox{E}\kern-.125emX}}
\begin{document}
\title{Designing two-dimensional limit-cycle oscillators with prescribed trajectories and phase-response characteristics}
    
\author{Norihisa Namura}
\thanks{Corresponding author. E-mail: namura.n.aa@m.titech.ac.jp}
\affiliation{Department of Systems and Control Engineering, Tokyo Institute of Technology, Tokyo 152-8552, Japan}

\author{Tsubasa Ishii}
\affiliation{Department of Systems and Control Engineering, Tokyo Institute of Technology, Tokyo 152-8552, Japan}

\author{Hiroya Nakao}
\affiliation{Department of Systems and Control Engineering, Tokyo Institute of Technology, Tokyo 152-8552, Japan}

\date{\today}


    \begin{abstract}
        We propose a method for designing two-dimensional limit-cycle oscillators with prescribed periodic trajectories and phase response properties based on the phase reduction theory,
        which gives a concise description of weakly-perturbed limit-cycle oscillators and is widely used in the analysis of synchronization dynamics.
        We develop an algorithm for designing the vector field with a stable limit cycle, which possesses a given shape and also a given phase sensitivity function.
        The vector field of the limit-cycle oscillator is approximated by polynomials whose coefficients are estimated by convex optimization.
        Linear stability of the limit cycle is ensured by introducing an upper bound to the Floquet exponent.
        The validity of the proposed method is verified numerically by designing several types of two-dimensional existing and artificial oscillators.
        As applications, we first design a limit-cycle oscillator with an artificial star-shaped periodic trajectory and demonstrate global entrainment.
        We then design a limit-cycle oscillator with an artificial high-harmonic phase sensitivity function
        and demonstrate multistable entrainment caused by a high-frequency periodic input.
    \end{abstract}

    \maketitle

    \section{INTRODUCTION}

    \label{sec:introduction}
        Synchronization of rhythmic systems have found wide applications in various fields of engineering in recent years. 
        Some examples include human-robot interactions~\cite{mortl2014rhythm,Jouaiti2018hebbian}, 
        frequency tuning or stabilization in electrical oscillators~\cite{millimeter1,millimeter2,lockrange1},
        power networks~\cite{Dorfler2012synchronization},
        suppression of pulsus alternans in the heart~\cite{wilson2017spatiotemporal,Monga2019optimal},
        and adjustment of circadian rhythms in shift works~\cite{Stone2019application}.
        Various methods for controlling
        synchronization dynamics of rhythmic systems have also been developed~\cite{Bai2019asymptotic,Oviedo2018synchronization}.
        
        Rhythmic systems are commonly modeled as limit-cycle oscillators, namely, nonlinear dynamical systems with stable limit-cycle trajectories~\cite{Strogatz}. 
        Some examples include
        brain waves~\cite{stankovski2015coupling,stankovski2017neural}, animal gaits~\cite{gait1,gait2,kobayashi2016,funato2016evaluation}, heartbeats and breathing~\cite{heart}, and passive walking~\cite{hobbelen2007limit,garcia1998simplest}.
        Synchronization of limit-cycle oscillators typically occur when they are periodically perturbed or mutually coupled.
        For example, when a periodic input is given to a limit-cycle oscillator, \textit{entrainment} or \textit{phase locking}~\cite{Kuramoto1984,pikovsky2001synchronization}, where the oscillator synchronizes with the periodic input, can be observed.
        When two or more oscillators are coupled together, mutual synchronization can be observed, in which their rhythms are aligned with each other.
        In the case of living organisms, the circadian rhythms, internal body clocks that exist in many organisms from bacteria to mammals, are entrained to the environmental light cycles.
        Fireflies flicker synchronously in unison due to the mutual interaction, namely, by sensing each other's luminescence.
        
        The phase reduction theory~\cite{Kuramoto1984,Hoppensteadt1997,Winfree2001,Ermentrout2010,Nakao2016,monga2019phase,Kuramoto2019,Ermentrout2019}
        is useful for analyzing synchronization dynamics of limit-cycle oscillators subjected to weak inputs.
        It represents the multidimensional state of a limit-cycle oscillator by using a single phase value defined along its limit cycle and approximately describes the oscillator dynamics by a one-dimensional phase equation.
        The phase equation, determined by the natural frequency and phase-response characteristics of the oscillator,
        has been extensively used for analyzing various types of synchronization dynamics of limit-cycle oscillators~\cite{Kuramoto1984,Winfree2001,Nakao2016}.
        In particular, various methods for optimizing or designing the periodic waveforms of the external inputs to achieve desired synchronization dynamics have been proposed 
        on the basis of the phase reduction theory~\cite{Harada2010optimal,Zlotnik2013optimal,Tanaka2014optimal,Tanaka2015optimal,Pikovsky2015maximizing,Zlotnik2016phase,Qiao2017entrainment,Kato2021optimization,Takata2021fast}.

        Studies on the design of oscillators with desirable periodic trajectories have also been conducted~\cite{righetti2009adaptive,isjpeert2013dynamical,ajalooeian2013general,Khoramshahi2017adaptive,hongu2022nonlinear}.
        For instance, rhythmic dynamics have been designed by constructing central pattern generators
        in the field of robotics~\cite{Ijspeert2001connectionist, Righetti2006programmable,ajalooeian2012design,Jouaiti2018hebbian,pasandi2022integrated}.
        Oscillatory trajectories have also been designed by using neural networks~\cite{Ruiz1998existence,Townley2000existence,Zegars2003trajectory,kuroe2005method,Jouffroy2008design}.
        Several studies have proposed methods to design dynamical systems with stable limit cycles of given shapes by constructing the vector fields~\cite{Okada2002Polynomial,okada2003hierachical,pasandi2019data,pasandi2020programmable}. 
        In applications, it is necessary to control the synchronization dynamics of the oscillators by appropriately choosing their phase-response characteristics and the applied external inputs or mutual interactions.
        It is also important to ensure the stability of the periodic trajectory.
        However, the phase-response characteristics of the oscillators, which are essentially important in determining their synchronization properties, have not been considered in most of the conventional methods.

        In this study, we propose a method for designing two-dimensional limit-cycle oscillators with given phase-response characteristics in addition to given stable periodic trajectories from the viewpoint of synchronization control.
        To ensure the stability of the limit cycle, we use the Floquet theory~\cite{hartman1964ordinary,guckenheimer1983nonlinear} and introduce an inequality constraint on the Floquet exponent characterizing the linear stability of the limit-cycle trajectory.
        We approximate the vector field of the oscillator by polynomials and optimize their coefficients to satisfy the given conditions.
        Our optimization problem is convex and can be numerically solved without difficulty to yield the target vector field.
        As applications, we design an oscillator with an artificial star-shaped periodic trajectory and an oscillator with an artificial phase sensitivity function that leads to multistable entrainment with multiple stable phase differences.
         
        This paper is organized as follows.
        We first explain the phase reduction theory and synchronization in Sec.~\ref{sec:theory}, then describe the method for designing the vector field of two-dimensional oscillators in Sec.~\ref{sec:method}.
        In Sec.~\ref{sec:results}, we verify the proposed method by numerical simulations for several types of two-dimensional oscillators.
        We conclude this study in Sec.~\ref{sec:conclusion}.


    \section{PHASE REDUCTION AND SYNCHRONIZATION}
    \label{sec:theory}
        In this section, we briefly explain {the} phase reduction theory and synchronization~\cite{Kuramoto1984,Hoppensteadt1997,Winfree2001,Ermentrout2010,Nakao2016,monga2019phase,Kuramoto2019,Ermentrout2019}.

        \subsection{Phase reduction theory}

            \subsubsection{Asymptotic phase}
                We consider a limit-cycle oscillator described by
                \begin{align}
                    \label{eq:system}
                    \ddt{} \bX(t) = \bF(\bX(t)),
                \end{align}
                where $\bX(t) \in \mathbb{R}^{N}$ is the system state at time $t$. 
                We assume that the system has an exponentially stable limit-cycle trajectory $\bX_{0}(t)$ with a natural period $T$ and frequency $\omega = 2\pi/T$, which is a $T$-periodic function of $t$ satisfying $\bX_{0}(t + T) = \bX_{0}(t)$. 
                
                First, we assign a phase $\theta \in [0, 2\pi)$ for each state on the limit cycle, where $0$ and $2\pi$ are considered identical,
                by choosing a state $\bX_{0}(0)$ on the limit cycle as the phase origin, $\theta=0$, and defining the phase of the state $\bX_{0}(t)$ at $t>0$ as $\theta = \omega t \mod{2\pi}$. 
                We denote the state on the limit cycle with phase $\theta$ by $\bcalX_{0}(\theta) = \bX_0(\theta/\omega)$.
            
                Next, we extend the definition of the phase to the basin of the limit cycle.
                The phase of a system state $\bX_{A}$ in the basin is defined as $\theta$
                if $\lim_{t \to \infty} \left\| \flow{t}{t_{0}}\bX_{A} - \flow{t}{t_{0}}\bcalX_{0}(\theta) \right\| = 0$ holds, 
                where $\flow{t}{t_{0}}$ is the flow of Eq.~\eqref{eq:system} and $\| \cdot \|$ is the $L_2$ norm. 
                That is, if the state $\flow{t}{t_{0}} \bX_{A}$ started from $\bX_{A}$ at $t_0$ converges to the state $\flow{t}{t_{0}} \bcalX_{0}(\theta)$ started from $\bcalX_{0}(\theta)$ at $t_0$ as $t \to \infty$, we consider that the state $\bX_{A}$ has the same phase $\theta$ as $\bcalX_{0}(\theta)$.
                This defines a phase function $\Theta$ for all states $\bX$ in the basin, which assigns a phase value $\theta(t) = \Theta(\bX(t))$ to the state $\bX(t)$ and satisfies
                \begin{align}
                    \label{eq:omega}
                    \ddt{} \theta(t) = \ddt{} \Theta(\bX(t)) = \omega.
                \end{align}
                The phase defined in this way is called the asymptotic phase,
                and the level sets of the phase function are called ``isochrons''~\cite{Kuramoto1984,Hoppensteadt1997,Winfree2001,Ermentrout2010,Nakao2016}.

            \subsubsection{Phase sensitivity function}

                The phase sensitivity function (PSF, also known as the infinitesimal phase resetting curve, iPRC) is a key quantity in analyzing synchronization of limit-cycle oscillators by weak inputs.
                It is defined by the gradient of the phase function at $\bcalX_{0}(\theta)$ on the limit cycle as 
                \begin{align}
                    \bZ(\theta) = \nabla \Theta(\bX)|_{\bX = \bcalX_{0}(\theta)}.
                \end{align}
                The PSF describes linear phase-response characteristics of the oscillator caused by weak external inputs~\cite{Kuramoto1984,Winfree2001,Brown2004,Ermentrout2010}.
            
                The PSF cannot be obtained analytically in general, so it should be calculated numerically or measured experimentally.
                If the mathematical model of the oscillator is known, the PSF $\bZ(\theta)$ can be obtained by numerically calculating the $2\pi$-periodic solutions to the following adjoint equation~\cite{Brown2004,Ermentrout2010,Kuramoto2019}:
                \begin{align}
                    \label{eq:adjZ}
                    \omega \ddtheta \bZ(\theta) &= - \tilde{\bJ}^{\top}(\theta) \bZ(\theta),
                \end{align}
                where $\tilde{\bJ}^{\top}(\theta)$ denotes the transposed Jacobian matrix of the vector field $\bF$ at $\bX = \bcalX_{0}(\theta)$.
                The PSF $\bZ(\theta)$ should satisfy $\bZ(\theta) \cdot \bF \left(\bcalX_{0}(\theta) \right) = \omega$ as a normalization condition.

                In this study, we consider the PSF as a function with a time argument,
                \begin{align}
                    \bZ(\theta) = \bZ(\omega t) \coloneqq \tilde{\bZ}(t).
                \end{align}
                For this $\tilde{\bZ}(t)$, the adjoint equation~\eqref{eq:adj_Zt} and normalization condition~\eqref{eq:norm_Zt} are given by
                \begin{align}
                    \label{eq:adj_Zt}
                    &\ddt{} \tilde{\bZ}(t) = -\bJ^{\top}(t) \tilde{\bZ}(t), \\
                    \label{eq:norm_Zt}
                    &\tilde{\bZ}(t) \cdot \bF \left(\bcalX_0(\theta(t)) \right) = \omega,
                \end{align}
                where $\bJ^{\top}(t) = \tilde{\bJ}^{\top}(\theta(t))$ is the transposed Jacobian matrix of $\bF$ evaluated at $\bcalX_{0}(\theta(t))$ on the limit cycle.

            \subsubsection{Phase reduction}

                Weakly-perturbed limit-cycle oscillators can generally be reduced to a one-dimensional phase equation by using the PSF $\bZ(\theta)$.
                Let us consider the dynamics of a weakly-perturbed limit-cycle oscillator described by
                \begin{align}
                    \label{eq:weakly_perturbed}
                    \ddt{} \bX(t) = \bF(\bX(t)) + \varepsilon \tilde{\BSeta}(\bX(t),t),
                \end{align}
                where $\varepsilon \tilde{\BSeta}(\bX,t)$ denotes the weak input and $0 < \varepsilon \ll 1$ is a small parameter representing the intensity of the input.
                Since $\varepsilon$ is small, the deviation of the oscillator state $\bX(t)$ from the state $\bcalX_{0}(\theta(t)) = \bX_{0}(t)$ on the limit cycle with the same phase $\theta(t)$ is $O(\varepsilon)$, i.e., $\bX(t) = \bcalX_{0}(\theta(t)) + O(\varepsilon)$.
                The time evolution of the phase $\theta(t) = \Theta(\bX(t))$ of the limit-cycle oscillator is then described by
                \begin{align}
                    \begin{aligned}
                        \ddt{} \theta(t) &= \ddt{} \Theta(\bX(t)) \\
                        &= \omega + \varepsilon\nabla\Theta(\bX(t)) \cdot \tilde{\BSeta}(\bX(t),t) \\
                        &= \omega + \varepsilon\nabla\Theta(\bcalX_{0}(\theta(t))) \cdot \tilde{\BSeta}(\bcalX_{0}(\theta(t)),t) + O(\varepsilon^{2}) \\
                        &= \omega + \varepsilon\bZ(\theta(t)) \cdot \BSeta(\theta(t),t) + O(\varepsilon^{2}).
                    \end{aligned}
                \end{align}
                
                Therefore, the phase $\theta(t)$ obeys the following approximate phase equation up to $O(\varepsilon)$:
                \begin{align}
                    \label{eq:phase_eq}
                    \ddt{} \theta(t) = \omega + \varepsilon\bZ(\theta(t)) \cdot \BSeta(\theta(t),t).
                \end{align}
                The simplicity of this phase equation has facilitated detailed and extensive analysis of synchronization caused by weak inputs~\cite{Kuramoto1984,Winfree2001,Brown2004,Ermentrout2010}.

        \subsection{Synchronization}
            By using phase reduction, we can analyze the synchronization of limit-cycle oscillators with a weak periodic external input. 
            We consider a limit-cycle oscillator subjected to weak periodic external input described by
            \begin{align}
                \ddt{} \bX(t) = \bF(\bX(t)) + \varepsilon \tilde{\bq}(\bX,t),
            \end{align}
            where $\tilde{\bq}(\bX,t)$ denotes a weak periodic external input with the period $\tau$ and the frequency $\Omega = 2\pi/\tau$ satisfying $\tilde{\bq}(\bX,t+\tau) = \tilde{\bq}(\bX,t)$.
            When the intensity $\varepsilon$ is sufficiently small, the time evolution of the phase $\theta(t)$ is given by
            \begin{align}
                \ddt{} \theta(t) = \omega + \varepsilon \bZ(\theta(t)) \cdot \bq(\theta(t),t)
            \end{align}
            by phase reduction, where the external input is approximately evaluated on the limit cycle, namely, $\bq(\theta(t), t) = \bq(\bcalX(\theta(t)), t)$.

            Assuming that the natural frequency $\omega$ of the oscillator and the frequency $\Omega$ of the periodic external input are sufficiently close, 
            we denote the frequency difference by $\varepsilon \Delta = \omega - \Omega$, where ``$\varepsilon \Delta$'' indicates that the frequency difference is $O(\varepsilon)$.
            The oscillator phase relative to the external input, $\phi(t) = \theta(t) - \Omega t$, obeys 
            \begin{align}
                \ddt{} {\phi(t)} = \varepsilon \left[ \Delta + \bZ(\phi{(t)} + \Omega t) \cdot \bq(\phi{(t)} + \Omega t,t) \right],
            \end{align}
            where the range of $\phi$ is extended outside $[0,2\pi)$ and $\bZ$ and $\bq$ are regarded as periodic functions of period $2\pi$.
            Since the right-hand side is $O(\varepsilon)$ and $\phi$ is a slow variable, this equation can further be simplified by the averaging approximation~\cite{Kuramoto1984,Hoppensteadt1997,Nakao2016}, 
            namely, by averaging the right-hand side over one period of the external input while keeping $\phi(t)$ fixed.
            This yields a single-variable autonomous system,
            \begin{align}
                \label{eq:phi_evol}
                \ddt{} {\phi(t)} = \varepsilon \left[ \Delta + \Gamma(\phi{(t)}) \right],
            \end{align}
            where we defined a $2\pi$-periodic phase coupling function
            \begin{align}
                \begin{aligned}
                    \Gamma(\phi) &= \frac{1}{\tau} \int_{0}^{\tau} \bZ(\phi + \Omega t) \cdot \bq(\phi + \Omega t,t) dt \\
                    &= \frac{1}{2\pi} \int_{0}^{2\pi} \bZ(\phi + \psi) \cdot \bq \left( \phi + \psi,\psi/\Omega \right) d\psi,
                \end{aligned}
            \end{align}
            where $\psi = \Omega t$ is the phase of the external input.

            We can analyze the entrainment of the oscillator to the weak periodic input by using this simplified equation.
            Specifically, the stable fixed point of Eq.~(\ref{eq:phi_evol}) corresponds to a phase-locking point, where the oscillator phase is entrained to the periodic input.

    
    \section{DESIGN OF OSCILLATOR DYNAMICS} 
    \label{sec:method}

		In this section, we propose a method for designing a stable two-dimensional limit-cycle oscillator with a prescribed periodic trajectory and PSF.

		\subsection{Conditions for the periodic trajectory and PSF}

			Our aim is to design a vector field with a prescribed stable limit cycle $\bp(t)$ and PSF $\tilde{\bZ}(t)$ of period $T$ and frequency $\omega = 2\pi / T$. 
			It is noted that the periodic trajectory and the PSFs are not completely independent.
            Specifically, the PSFs should satisfy the normalization condition $\dot{\bp}(t)\cdot \tilde{\bZ}(t) = \omega$.
            
            We approximate the vector field of the oscillator by using polynomials as~\cite{namura2022estimating}
            \begin{align}
                \bF(\bX) = 
                \begin{bmatrix}
                    F_{1}(\bX) \\ F_{2}(\bX)
                \end{bmatrix}
                \simeq
                \begin{bmatrix}
                    \bU^{\top}(\bX)\bzeta_{1} \\ \bU^{\top}(\bX)\bzeta_{2}
                \end{bmatrix}
                ,
            \end{align}
            where $\bX = [x_1\ x_2]^{\top}$,
            \begin{align}
                \bU(\bX) = 
                \begin{bmatrix}
                    1 & \overline{x_{1}} & \overline{x_{2}} & \overline{x_{1}^{2}} & \overline{x_{1}x_{2}} & \overline{x_{2}^{2}} & \cdots & \overline{x_{2}^{n}}
                \end{bmatrix}
                ^{\top} \in \mathbb{R}^{P}
            \end{align}
            is a vector of monomials, $n$ is the maximum degree of the polynomial, and $\bzeta_{1}, \bzeta_{2} \in \mathbb{R}^{P}$ are the coefficient vectors.
            Here, the overline denotes standardization ($\overline{z} \coloneqq (z - \mu)/\sigma$), where $\mu$ and $\sigma$ are the mean and the standard deviation of $z$, respectively.
            
            We assume that the system has a non-intersecting differentiable periodic trajectory $\bp(t) = [p_1(t)\ p_2(t)]^{\top}$ and impose the following condition on the vector field:
            \begin{align}
                \label{eq:periodic}
                \bF(\bp(t)) = \ddt{} \bp(t).
            \end{align}
            This equation can be expressed in the polynomial approximation as follows:
            \begin{align}
                \begin{aligned}
                    \bU^{\top}(\bp(t)) \bzeta_{1} \simeq \ddt{} p_{1}(t), \\
                    \bU^{\top}(\bp(t)) \bzeta_{2} \simeq \ddt{} p_{2}(t).
                \end{aligned}
            \end{align}

            Introducing the coefficient $\bxi = \left[ \bzeta_{1}^{\top}\ \bzeta_{2}^{\top} \right]^{\top} \in \mathbb{R}^{2P}$ and 
            discretizing the time $t$ into $L$ points, $t_{k} = (k-1)\Delta t,\; (k = 1, \dots, L)$ with $\Delta t = T/L$, 
            we express the left-hand side as
            \begin{align}
                \begin{bmatrix}
                    \bU^{\top}(\bp(t_{k})) & \bm{0} \\
                    \bm{0} & \bU^{\top}(\bp(t_{k}))
                \end{bmatrix}
                \bxi \coloneqq \bA_{\bp,k} \bxi 
            \end{align}
            for each $t_{k}$, where $\bA_{\bp,k} \in \mathbb{R}^{2 \times 2P}$.
            We also denote the right-hand side by
            \begin{align}
                \bb_{\bp,k} \coloneqq
                \begin{bmatrix}
                    \dot{p}_{1}(t_{k}) \\ \dot{p}_{2}(t_{k})
                \end{bmatrix}
                ,
            \end{align}
            where $\bb_{\bp,k} \in \mathbb{R}^{2}$.
            The difference between both sides of Eq.~(\ref{eq:periodic}) under polynomial approximation is then expressed as
            \begin{align}
                \label{eq:error_p}
                \bA_{\bp,k} \bxi - \bb_{\bp,k}.
            \end{align}

            Next, we introduce the conditions for the PSF into the vector field.
            Assuming that the oscillator has a PSF $\tilde{\bZ}(t) = [ \tilde{Z}_{1}(t)\ \tilde{Z}_{2}(t) ]^{\top}$, we require the adjoint equation~\eqref{eq:adj_Zt}, represented as a function of $t$, is satisfied.
            This equation is expressed in the polynomial approximation as
            \begin{align}
                \begin{aligned}
                    \tilde{Z}_{1}(t) \bU_{1}^{\top}(\bp(t))\bzeta_{1} + \tilde{Z}_{2}(t) \bU_{1}^{\top}(\bp(t))\bzeta_{2} \simeq -\ddt{} \tilde{Z}_{1}(t), \\
                    \tilde{Z}_{1}(t) \bU_{2}^{\top}(\bp(t))\bzeta_{1} + \tilde{Z}_{2}(t) \bU_{2}^{\top}(\bp(t))\bzeta_{2} \simeq -\ddt{} \tilde{Z}_{2}(t),
                \end{aligned}
            \end{align}
            where $\bU_{i}(\bX) = \left( \nabla \bU(\bX) \right)_{i}$.

            Introducing the coefficient $\bxi$ and $\{ t_{k} \}_{k=1}^{L}$,
            we express the left-hand side as
            \begin{align}
                \begin{bmatrix}
                    \tilde{Z}_{1}(t_{k}) \bU_{1}^{\top}(\bp(t_{k})) & \tilde{Z}_{2}(t_{k}) \bU_{1}^{\top}(\bp(t_{k})) \\
                    \tilde{Z}_{1}(t_{k}) \bU_{2}^{\top}(\bp(t_{k})) & \tilde{Z}_{2}(t_{k}) \bU_{2}^{\top}(\bp(t_{k}))
                \end{bmatrix}
                \bxi \coloneqq \bA_{\bZ,k} \bxi
            \end{align}
            for each $t_{k}$, where $\bA_{\bZ,k} \in \mathbb{R}^{2 \times 2P}$,
            and also denote the right-hand side as
            \begin{align}
                \bb_{\bZ,k} \coloneqq
                \begin{bmatrix}
                    -\dot{\tilde{Z}}_{1}(t_{k}) \\ -\dot{\tilde{Z}}_{2}(t_{k})
                \end{bmatrix}
                ,
            \end{align}
            where $\bb_{\bZ,k} \in \mathbb{R}^{2}$.
            The difference between both sides of Eq.~\eqref{eq:adj_Zt} under polynomial approximation is then expressed as
            \begin{align}
                \label{eq:error_Z}
                \bA_{\bZ,k} \bxi - \bb_{\bZ,k}.
            \end{align}

            We seek the polynomial vector field that satisfies Eq.~\eqref{eq:periodic} and~\eqref{eq:adj_Zt} as much as possible
            by minimizing the overall approximation errors for the periodic trajectory and PSF over one period $T$. 
            By introducing matrices
            \begin{align}
                \bA &= 
                \begin{bmatrix}
                    \bA_{\bp,1}^{\top} & \cdots & \bA_{\bp,L}^{\top} & \bA_{\bZ,1}^{\top} & \cdots & \bA_{\bZ,L}^{\top}
                \end{bmatrix}
                ^{\top} \in \mathbb{R}^{4L \times 2P}
            \end{align}
            and
            \begin{align}
                \bb &= 
                \begin{bmatrix}
                    \bb_{\bp,1}^{\top} & \cdots & \bb_{\bp,L}^{\top} & \bb_{\bZ,1}^{\top} & \cdots & \bb_{\bZ,L}^{\top}
                \end{bmatrix}
                ^{\top} \in \mathbb{R}^{4L},
            \end{align}
            the sum of the squared errors in Eqs.~\eqref{eq:error_p} and~\eqref{eq:error_Z} can be expressed as
            \begin{align}
                E = \frac{1}{2} \| \bA\bxi - \bb \|^{2}.
                \label{eq:error_sum}
            \end{align}
            This error is the objective function to be minimized in our method.

            In addition to the above conditions, we introduce regularization 
            of the vector field by adding the squared norm $\| \bxi \|^2$ of the coefficient $\bxi$ with an appropriate weight $\gamma$ to the objective function in order to obtain a better model.
            This regularization prevents the coefficients of the vector field from becoming excessively large, 
            which is expected to reduce the complexity of the vector field and ensures the uniqueness of the solution of the optimization problem.
    
		\subsection{Condition for the linear stability}
        
            In addition to the conditions for the trajectory and PSF, we also impose the condition for the  stability of the periodic trajectory on the vector field.
            From the Floquet theory~\cite{hartman1964ordinary,guckenheimer1983nonlinear}, 
            the two Floquet exponents $\lambda_{1}$ and $\lambda_{2}$ characterizing the linear stability of the periodic trajectory satisfy
            \begin{align}
                \exp(\lambda_{1}T)\exp(\lambda_{2}T) = \exp\left( \int_{0}^{T} \tr(\bJ(t))dt \right),
            \end{align}
            where $\bJ(t)$ is the Jacobian matrix at $\bX_0(t)$ on the periodic trajectory and $\tr$ denotes the trace of a matrix.
            Taking the logarithm, we obtain
            \begin{align}
                \lambda_{1} + \lambda_{2} = \frac{1}{T} \int_{0}^{T} \tr(\bJ(t))dt
            \end{align}
            since $\lambda_{1}$ and $\lambda_{2}$ are real for two-dimensional limit-cycle oscillators.
            
            From the condition \eqref{eq:periodic} for the periodic trajectory, the first Floquet exponent $\lambda_{1} = 0$, thus we only need to evaluate $\lambda = \lambda_{2}$, which should be negative for the stability.
            Discretizing the time $t$ into $\{ t_{k} \}_{k=1}^{L}$, we can approximate the above integral as
            \begin{align}
                \lambda \simeq \frac{1}{L} \sum_{k=1}^{L} \left( \bJ_{11}(t_{k}) + \bJ_{22}(t_{k})\right),
            \end{align} 
            which is expressed in the polynomial approximation as
            \begin{align}
                \begin{aligned}
                    \lambda = \lambda_{2} &\simeq \frac{1}{L} \sum_{k=1}^{L} \left( \bU_{1}^{\top}(\bp(t_{k}))\bzeta_{1} + \bU_{2}^{\top}(\bp(t_{k}))\bzeta_{2} \right) \\
                    &= \frac{1}{L} \sum_{k=1}^{L}
                    \begin{bmatrix}
                        \bU_{1}^{\top}(\bp(t_{k})) & \bU_{2}^{\top}(\bp(t_{k}))
                    \end{bmatrix}
                    \bxi \coloneqq \bC \bxi.
                \end{aligned}
            \end{align} 
            We impose the condition
            \begin{align}
                \lambda = \bC\bxi \leq \lambda_{\mathrm{tol}}
            \end{align}
            for some maximum tolerance value $\lambda_{\mathrm{tol}} < 0$
            so that the periodic trajectory is sufficiently stable.

        \subsection{Optimization problem}

            Summarizing the conditions for the periodic trajectory~$\bp(t)$, PSF~$\tilde{\bZ}(t)$, and maximal tolerance $\lambda_{\mathrm{tol}}$ of the linear stability explained in the previous subsections, our optimization problem for the coefficient vector $\bxi$ is formulated as
            \begin{gather}
                \begin{gathered}
                    \bxi^{*} = \argmin_{\bxi} \frac{1}{2} \| \bA\bxi - \bb \|^{2} + \gamma \|\bxi\|^{2} \\
                    \mathrm{s.t.} \quad \bC \bxi \leq \lambda_{\mathrm{tol}}. 
                \end{gathered}
            \end{gather}
            Since this is a quadratic programming problem for the parameter $\bxi$ with a linear inequality constraint, which is convex, 
            it can be easily solved uniquely and globally.   

            In the numerical implementation, the weight parameter $\gamma$ is chosen 
            so that the designed oscillator has a stable limit cycle with the prescribed properties and no unnecessary attractors arises around the limit cycle.
            Also, to match the scales of the errors for the periodic trajectory and PSF in Eqs.~\eqref{eq:error_p} and~\eqref{eq:error_Z},
            the conditions for the PSF are multiplied by an appropriate normalizing constant.


    \section{RESULTS}
    \label{sec:results}
        \subsection{Reconstruction of existing oscillators}
        In this subsection, we test the validity of the proposed method by reconstructing the periodic trajectories and PSFs of two well-known oscillators.

            \subsubsection{van der Pol oscillator}

                As the first example, we design an oscillator with a periodic trajectory and PSF of the van der Pol~(vdP) oscillator~\cite{van1927vii,van1927frequency,van1926lxxxviii}.
                The vdP oscillator is described by
                \begin{align}
                    \label{eq:vdp}
                    \ddt{}
                    \begin{bmatrix}
                        x_{1} \\ x_{2}
                    \end{bmatrix}
                    = 
                    \begin{bmatrix}
                        x_{2} \\
                        \nu \left( 1 - x_{1}^{2} \right)x_{2} - x_{1}
                    \end{bmatrix}
                    ,
                \end{align}
                where we assume $\nu = 3$.
                This vdP oscillator has a period $T = 8.860$, natural frequency $\omega = 0.7092$, and the second Floquet exponent $\lambda_2 = -3.9396$. 
                The PSF can be calculated by numerically solving the adjoint equation~\eqref{eq:adjZ}.
				
                We designed the vector field 
                using the time interval $\Delta t = 0.005$, the degree of the polynomial as $n = 10$, the weight parameter $\gamma = 1.0$, and the maximum tolerance value of the second Floquet exponent $\lambda_{\mathrm{tol}} = -0.5$.
                By optimization, we obtained an oscillator with a period $T = 8.860$, natural frequency $\omega = 0.7092$, and the second Floquet exponent $\lambda_2 = -3.7260$.
                The period and natural frequency are almost identical to those of the original vdP oscillator,
                while the second Floquet exponent is much smaller than the assumed maximal tolerance value and close to the value of the original vdP oscillator.

\begin{figure}[t]
                    \centering
                    \includegraphics[width=\hsize]{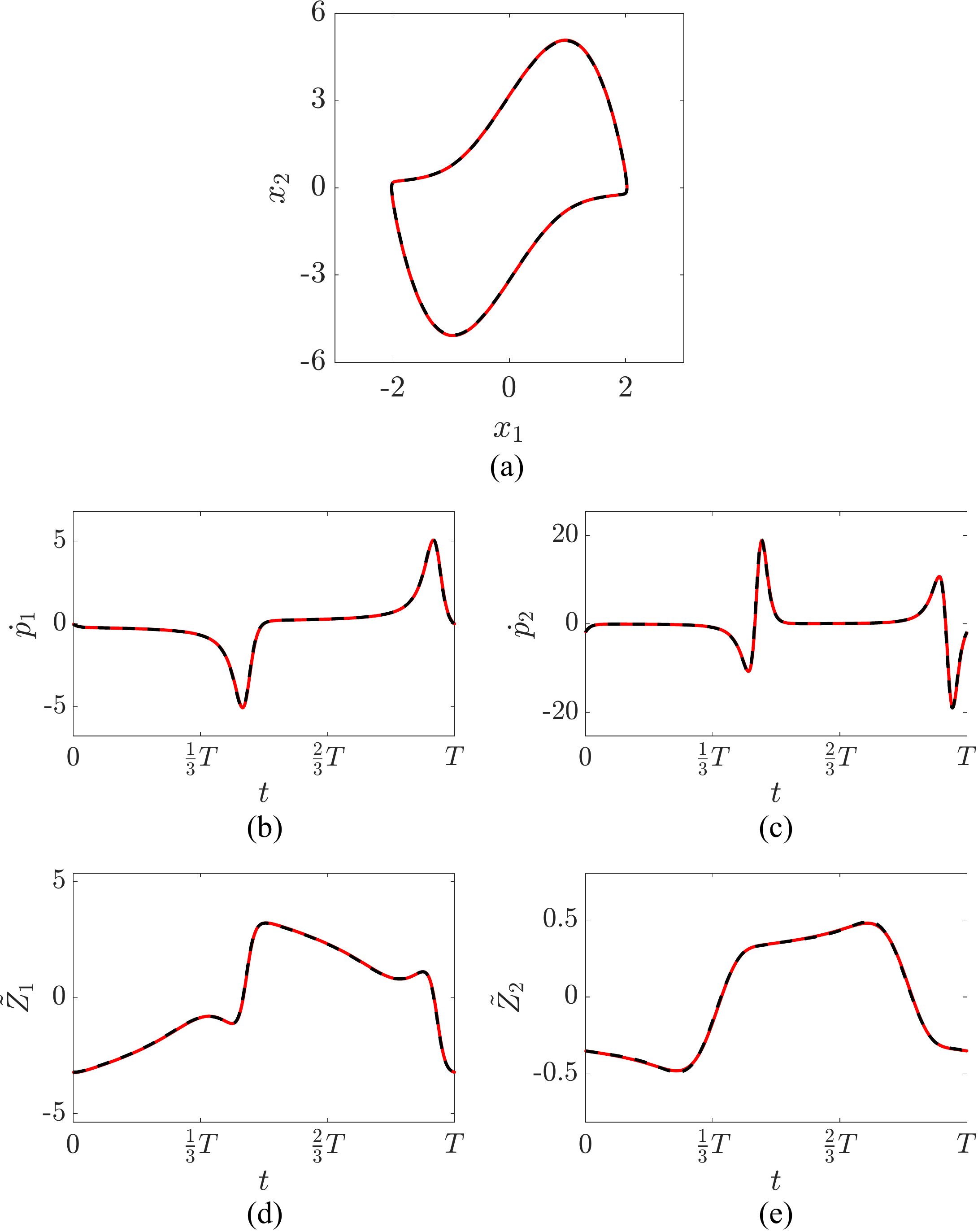}
                    \caption{
                        Periodic trajectories and PSFs of the designed oscillator and the original vdP oscillator.
                        (a)~Periodic trajectories.
                        (b,c)~Velocities on the periodic trajectories.
                        (b)~$x_{1}$ component, (c)~$x_{2}$ component.
                        (d,e)~PSFs.
                        (d)~$x_{1}$ component, (e)~$x_{2}$ component.
                        In each graph, the red line shows the designed functional form and the black dotted line shows the original one, respectively.
                    }
                    \label{fig:namur1}
                \end{figure}

\begin{figure}[t]
                    \centering
                    \includegraphics[width=0.9\hsize]{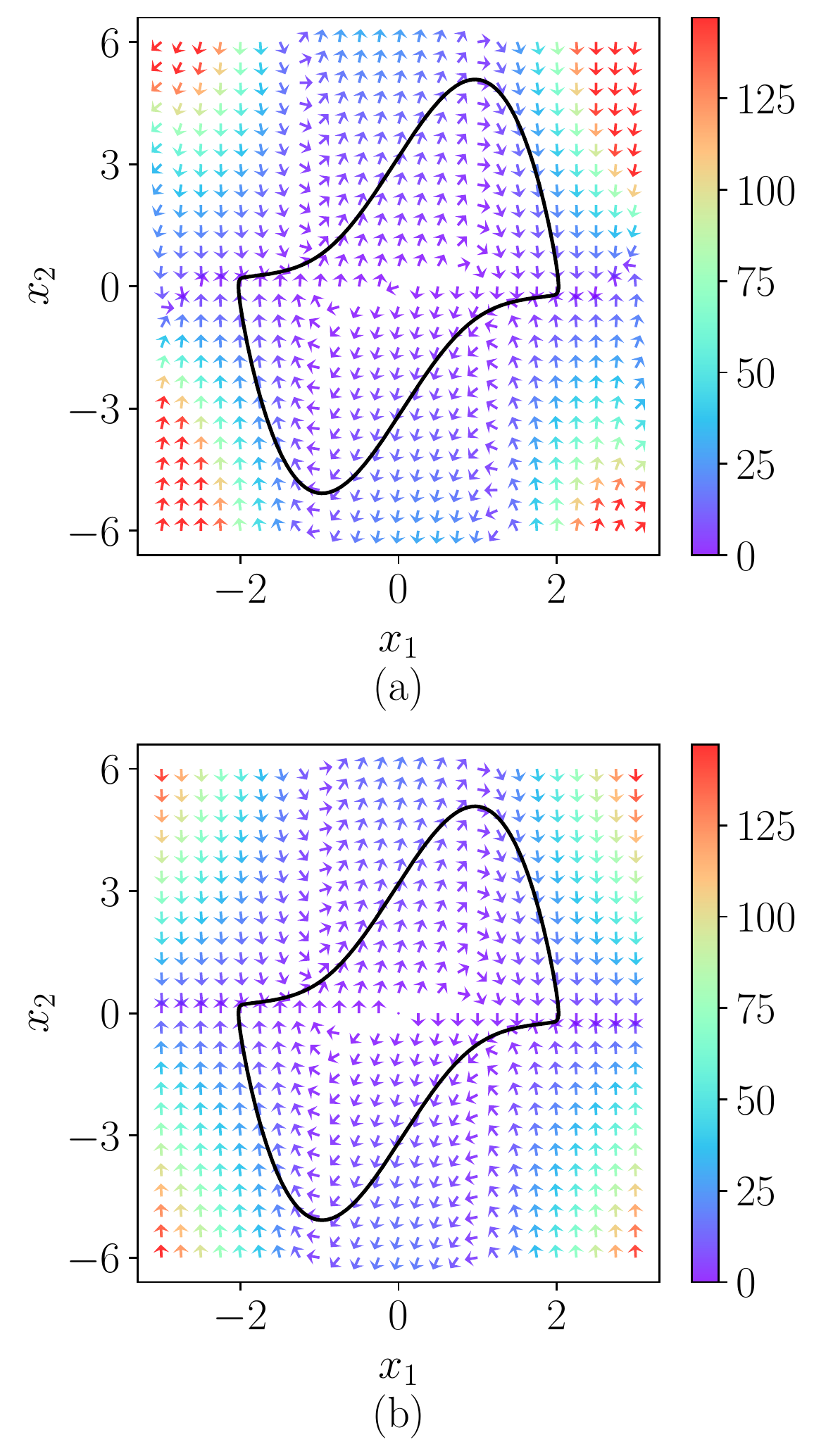}
                    \caption{
                        Vector fields of the designed oscillator and the original vdP oscillator.
                        (a)~designed, (b)~original.
                    }
                    \label{fig:namur2}
                \end{figure}

                The periodic trajectory on the $(x_1, x_2)$ plane and the velocity $\dot{p}_1(t)$ and $\dot{p}_2(t)$ of the designed oscillator are compared with those of the original vdP oscillator
                in Figs.~\ref{fig:namur1}~(a), (b), and (c);
                both of them are almost identical to those of the original oscillator.
                The vector field of the designed oscillator is compared with that of the original vdP oscillator in Fig.~\ref{fig:namur2}.
                Both vector fields are similar in the whole region, though the scales of the vector fields are slightly different.
                Figures~\ref{fig:namur1}~(d) and (e) show the PSFs of the designed oscillator and the original vdP oscillator, which are almost identical to each other.
                Thus, our method can reproduce the functional forms of the periodic orbit and PSF of the vdP oscillator successfully.
                

            \subsubsection{FitzHugh-Nagumo oscillator}
                As the second example, we design an oscillator with a periodic trajectory and PSF of the FitzHugh-Nagumo~(FHN) oscillator~\cite{Nagumo1962Active} described by
                \begin{align}
                    \ddt{}
                    \begin{bmatrix}
                        x_{1} \\ x_{2}
                    \end{bmatrix}
                    = 
                    \begin{bmatrix}
                        x_{1}(x_{1} - a)(1 - x_{1}) - x_{2} \\
                        c(x_{1} - bx_{2})
                    \end{bmatrix}
                    ,
                \end{align}
                where we consider $a = -0.1$, $b = 0.5$, and $c = 0.01$.
                The FHN oscillator is a fast-slow system with a large timescale difference between the two variables.
                This FHN oscillator has a period $T = 126.5$, natural frequency $\omega = 0.0497$, and the second Floquet exponent $\lambda_2 = -0.4586$.
                The PSF is calculated by solving the adjoint equation numerically.
                
                We designed the vector field 
                by setting the time interval as $\Delta t = 0.05$, the degree of the polynomial as $n = 10$, the weight parameter $\gamma = 1.0 \times 10^{-3}$, and the maximum tolerance value of the second Floquet exponent as $\lambda_{\mathrm{tol}} = -0.5$.
                The designed oscillator had a period $T = 126.5$, natural frequency $\omega = 0.0497$, and the second Floquet exponent $\lambda_2 = -0.5000$, 
                where the period and the natural frequency are almost identical to the original FHN oscillator.
                We note that the maximum tolerance value of the second Floquet exponent $\lambda_{\mathrm{tol}} = -0.5$ is below the original value $\lambda_2 = -0.4586$ of the FHN oscillator.                 
                This shows that the stability of the periodic trajectory can be varied from that of the original oscillator to a certain extent.
                
\begin{figure}[t]
                    \centering
                    \includegraphics[width=\hsize]{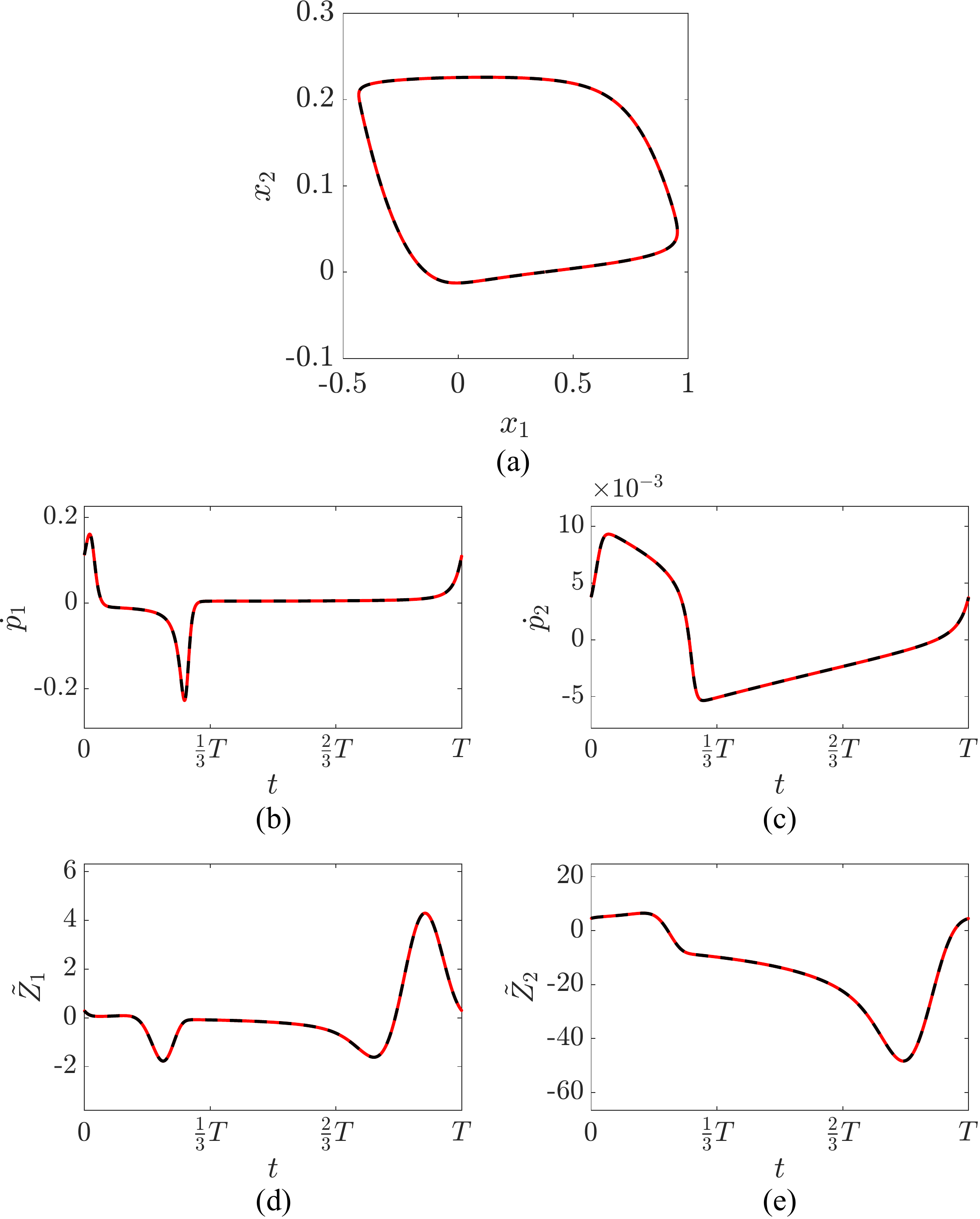}
                    \caption{
                        Periodic trajectories and PSFs of the designed oscillator and the original FHN oscillator.
                        (a)~Periodic trajectories.
                        (b,c)~Velocities on the periodic trajectories.
                        (b)~$x_{1}$ component, (c)~$x_{2}$ component.
                        (d,e)~PSFs.
                        (d)~$x_{1}$ component, (e)~$x_{2}$ component.
                        In each graph, the red line shows the designed functional form and the black dotted line shows the original one, respectively.
                    }
                    \label{fig:namur3}
                \end{figure}

\begin{figure}[t]
                    \centering
                    \includegraphics[width=0.9\hsize]{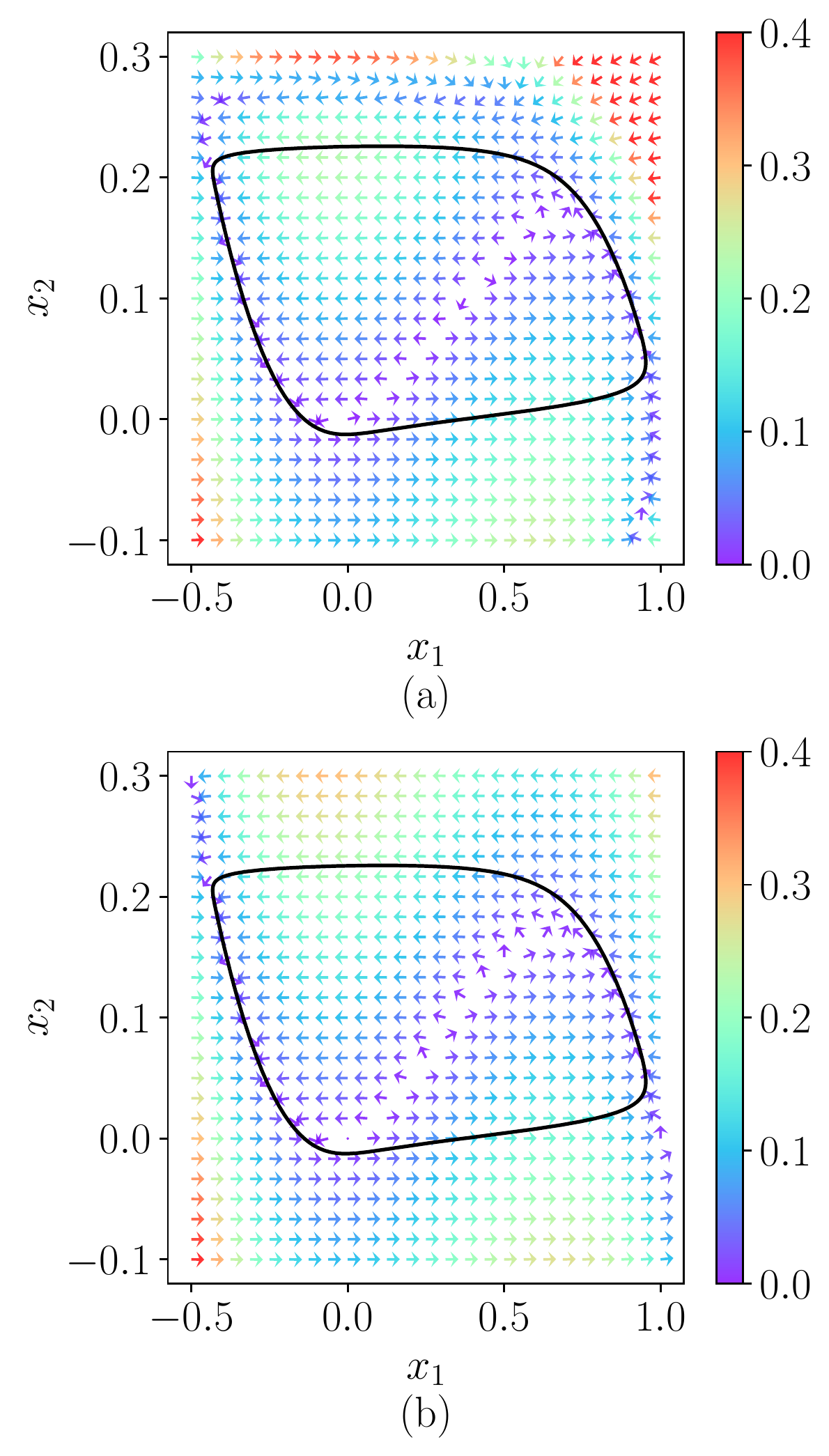}
                    \caption{
                        Vector fields of the designed oscillator and the original FHN oscillator.
                        (a)~designed, (b)~original.
                    }
                    \label{fig:namur4}
                \end{figure}

				As shown in Figs.~\ref{fig:namur3}~(a), (b), and (c), the original and designed oscillators have almost identical limit cycles.
                The vector field of the designed oscillator
                is compared to that of the original FHN oscillator
                in Fig.~\ref{fig:namur4}.
                In the vicinity of the periodic trajectory, the vector fields are almost similar and the nullclines of the fast component look alike. 
                They are considerably different in the regions far away from the periodic trajectory, though it does not affect the resulting limit cycles largely.
                The PSFs of the original and designed oscillators are also almost identical as shown in Figs.~\ref{fig:namur3}~(d) and (e).


        \subsection{Design of an oscillator with an artificial periodic trajectory and PSF}

            \subsubsection{Design of an oscillator with a star shape}

                Next, we design a limit-cycle oscillator with an artificial periodic trajectory and PSF.
                We assume that the oscillator has the following periodic trajectory~\eqref{eq:arti_period_trajectory2} and PSF~\eqref{eq:arti_psf2}:
                \begin{align}
                    \label{eq:arti_period_trajectory2}
                    \bp(t) &= 
                    \begin{bmatrix}
                        \sqrt{2}\cos(t) + \frac{1}{4}\sin(4t) \\ \sqrt{2}\sin(t) + \frac{1}{4}\cos(4t)
                    \end{bmatrix}
                    , \\ 
                    \label{eq:arti_psf2}
                    \tilde{\bZ}(t) &= 
                    \begin{bmatrix}
                        -\sqrt{2}\sin(t) - \cos(4t) \\ \sqrt{2}\cos(t) + \sin(4t)
                    \end{bmatrix}
                    ,
                \end{align}
                where the given PSF satisfies the normalization condition $\dot{\bp}(t) \cdot \tilde{\bZ}(t) = \omega = 1$.
                The periodic trajectory has a star shape with a $5$-fold symmetry, which is not often seen in real-world oscillators. The oscillation period is $T = 2\pi$.
                
                We designed the vector field that possesses the given artificial periodic trajectory and PSF
                by using the number of data points $L = 1000$, the degree of the polynomial $n = 10$, the weight parameter $\gamma = 1.0 \times 10^{-3}$, and the maximum tolerance value of the second Floquet exponent $\lambda_{\mathrm{tol}} = -1$.
                We obtained an oscillator with a period $T = 6.2832$, natural frequency $\omega = 1.0000$, which are almost identical to the assumed values, and the second Floquet exponent $\lambda_2 = -1.0001$, 
                which is almost equal to $\lambda_{\mathrm{tol}}$.
                
\begin{figure}[t]
                    \centering
                    \includegraphics[width=\hsize]{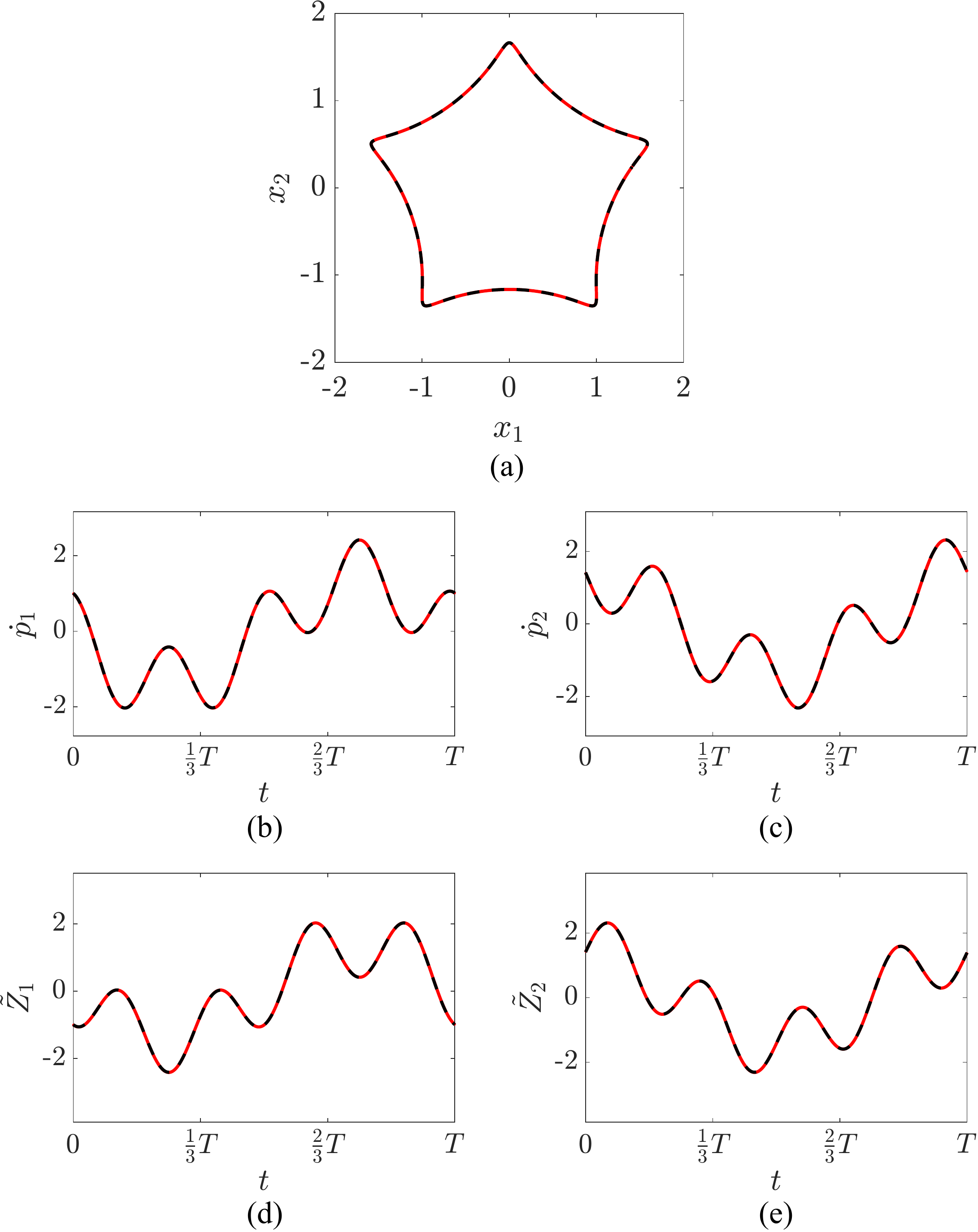}
                    \caption{
                        Periodic trajectory and PSFs of the designed oscillator and the prescribed functional forms.
                        (a)~Periodic trajectories.
                        (b,c)~Velocities on the periodic trajectories.
                        (b)~$x_{1}$ component, (c)~$x_{2}$ component.
                        (d,e)~PSFs.
                        (d)~$x_{1}$ component, (e)~$x_{2}$ component.
                        In each graph, the red line shows the designed functional form and the black dotted line shows the original one, respectively.
                    }
                    \label{fig:namur5}
                \end{figure}

\begin{figure}[t]
                    \centering
                    \includegraphics[width=0.9\hsize]{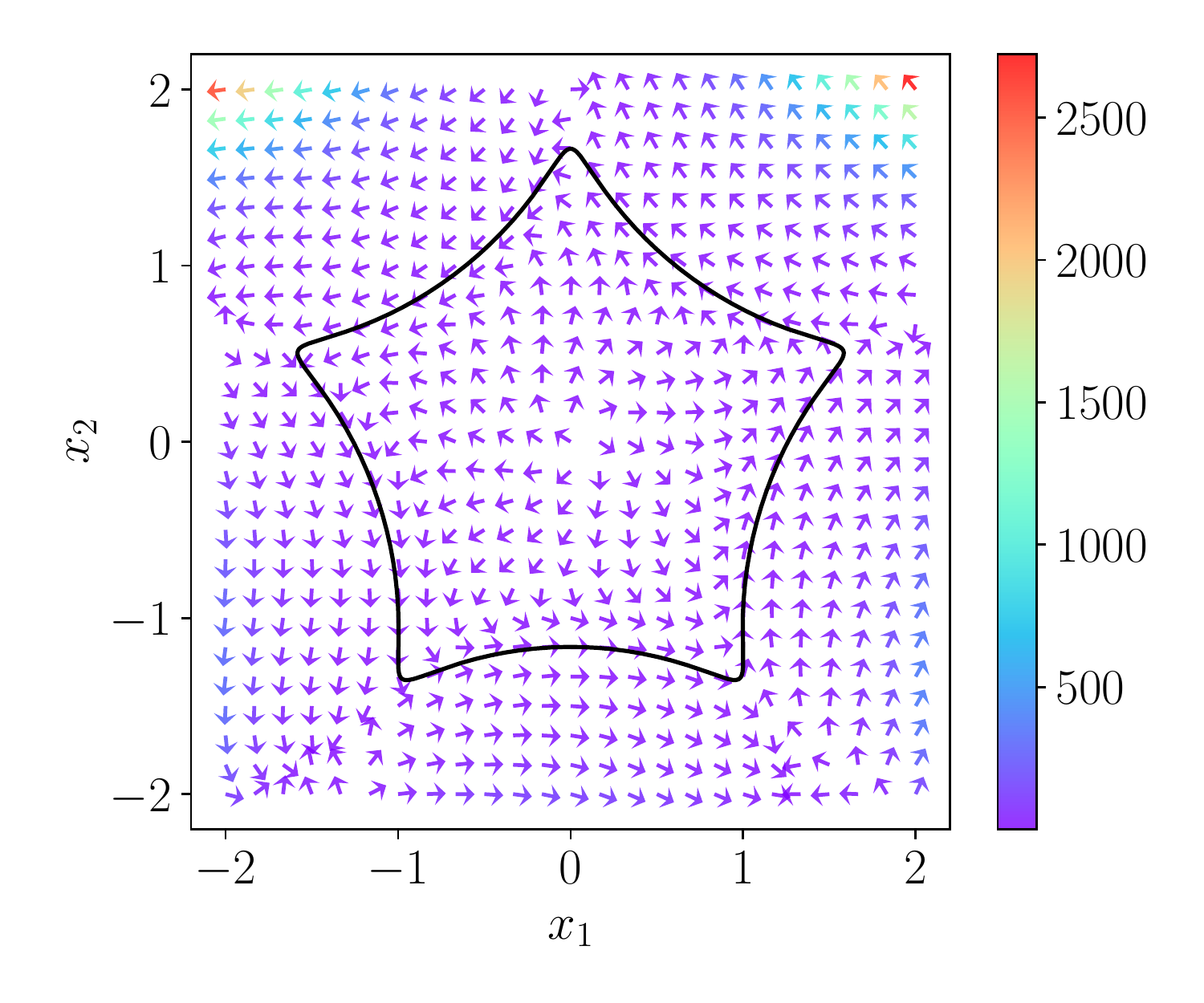}
                    \caption{Vector field and limit cycle of the designed oscillator with an artificial star-shaped periodic trajectory.}
                    \label{fig:namur6}
                \end{figure}

                The limit cycle of the designed oscillator and the given periodic trajectory are compared in Figs.~\ref{fig:namur5}~(a), (b), and (c).
                Both the periodic trajectory and the velocities on the periodic trajectory are almost identical to the assumed ones.
                The PSF of the designed oscillator is shown in Figs.~\ref{fig:namur5}~(d) and (e),
                which is also almost identical to the assumed one.
                The vector field of the designed oscillator is shown in Fig.~\ref{fig:namur6}.

            \subsubsection{Global entrainment}

\begin{figure}[t]
                    \centering
                    \includegraphics[width=0.75\hsize]{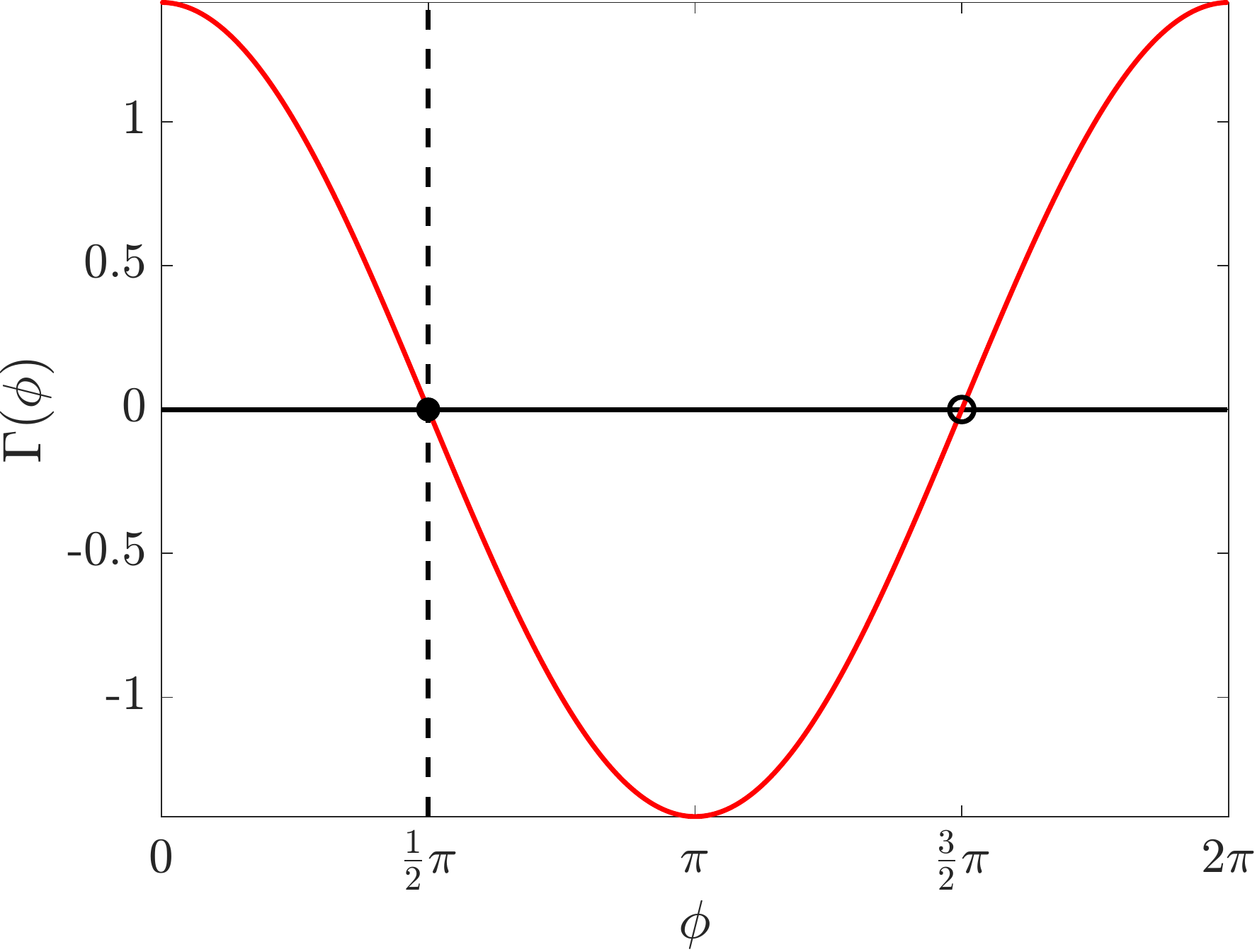}
                    \caption{
                        Phase coupling function $\Gamma(\phi)$ for global entrainment.
                            The black solid point shows the stable fixed point and the black circle point shows the unstable fixed point.
                            If $\dot{\phi} = 0$ and $\Gamma'(\phi) < 0$ at $\phi = \phi^{*}$, 
                            where $\Gamma'(\phi)$ denotes the derivative of $\Gamma$, $\phi^{*}$ is a stable fixed point.
                    }
                    \label{fig:namur7}
                \end{figure}

                Here, we demonstrate global entrainment of the star-shaped oscillator designed above. 
                We apply an external periodic input with a frequency $\omega$, which is equal to the natural frequency. 
                Assuming that the external periodic input has the form of $\bq(t) = \left[ -\sin(\omega t), \cos(\omega t) \right]^{\top}$, 
                we can write the reduced phase equation of the oscillator as
                \begin{align}
                    \ddt{} {\theta(t)} = \omega + \varepsilon \bZ(\theta{(t)}) \cdot \bq(t).
                \end{align}
                From Eq.~\eqref{eq:phi_evol}, the time evolution of the relative phase $\phi{(t)} = \theta(t) - \omega t$ obeys
                \begin{align}
                    \begin{aligned}
                        \ddt{} {\phi(t)} &= \varepsilon\Gamma(\phi{(t)}) \\
                        &= \sqrt{2}\varepsilon \cos(\phi{(t)}).
                    \end{aligned}
                \end{align}
                The phase coupling function $\Gamma(\phi)$ is shown in Fig.~\ref{fig:namur7}. 
                There are two fixed points (one stable and one unstable) satisfying $\dot{\phi} = 0$ within $\phi \in [0,2\pi)$;
                the only stable fixed point is at $\phi{^{*}} = \pi/2$, indicating that the relative phase converges to this point irrespective of the initial value (except for the unstable fixed point $\phi=3\pi/2$), namely, the oscillator exhibits global $1:1$ entrainment.

\begin{figure}[t]
                    \centering
                    \includegraphics[width=\hsize]{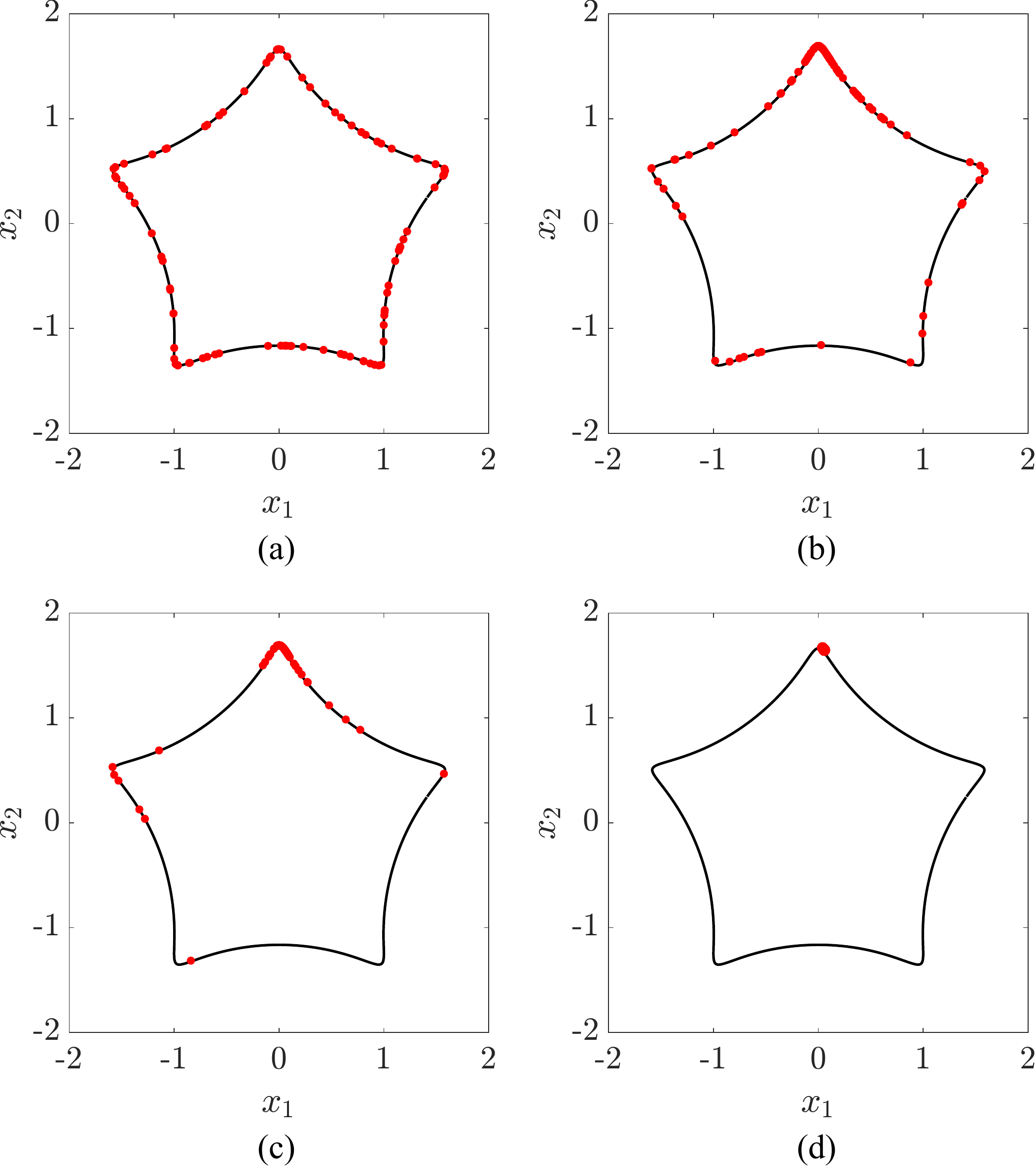}
                    \caption{
                        Global entrainment of $N=100$ designed artificial oscillators.
                        {(a)~$t = 0$, (b)~$t = 20\pi$, (c)~$t = 40\pi$, and (d)~$t = 100\pi$.}
                        In (d), we can see the convergence to the unique stable fixed point.
                    }
                    \label{fig:namur8}
                \end{figure}

                We performed numerical simulations to confirm that global entrainment occurs.
                We assumed $\varepsilon = 0.01$ and evolved $100$ independent oscillators from $100$ random initial points on the star-shape periodic trajectory.
                The oscillator states at time $t = 0,\; 20\pi,\; 40\pi$, and $100\pi$ are shown in Fig.~\ref{fig:namur8}.
                As time passes, the phases of the oscillators gradually form a cluster, and at $t = 100\pi$, 
                the oscillators almost converge to the unique fixed point.                 


        \subsection{Design of an oscillator exhibiting multistable entrainment}

            \subsubsection{Design of an oscillator with a high-harmonic PSF}

                Finally, we design an oscillator with an artificial PSF that leads to nontrivial multistable entrainment.
                We assume that the oscillator has the following periodic trajectory and PSF:
                \begin{align}
                    \label{eq:arti_period_trajectory1}
                    \bp(t) &= 
                    \begin{bmatrix}
                        \cos(t) \\ \sin(t)
                    \end{bmatrix}
                    , \\ 
                    \label{eq:arti_psf1}
                    \tilde{\bZ}(t) &= 
                    \begin{bmatrix}
                        -\sin(5t) \\ 2\cos(t) - 2\cos(3t) + \cos(5t)
                    \end{bmatrix}
                    ,
                \end{align}
                where the given PSF satisfies the normalization condition $\dot{\bp}(t) \cdot \tilde{\bZ}(t) = \omega = 1$.
                This periodic trajectory is simply a unit circle and the system state rotates with a constant frequency $\omega = 1$, that is, the period is $T = 2\pi$.
                
                We designed the vector field that possesses the given artificial periodic trajectory and PSF
                by setting the number of data points as $L = 1000$, the degree of the polynomial as {$n = 7$}, the weight parameter {$\gamma = 1.0 \times 10^{-2}$}, and the maximum tolerance value of the second Floquet exponent as $\lambda_{\mathrm{tol}} = -1$.
                We obtained an oscillator with a period $T = 6.2832$, natural frequency $\omega = 1.0000$, which are almost identical to the assumed values,
                and the second Floquet exponent $\lambda_2 = -0.9998$, 
                which is almost equal to the maximum tolerance value $\lambda_{\mathrm{tol}}$.
                
\begin{figure}[t]
                    \centering
                    \includegraphics[width=\hsize]{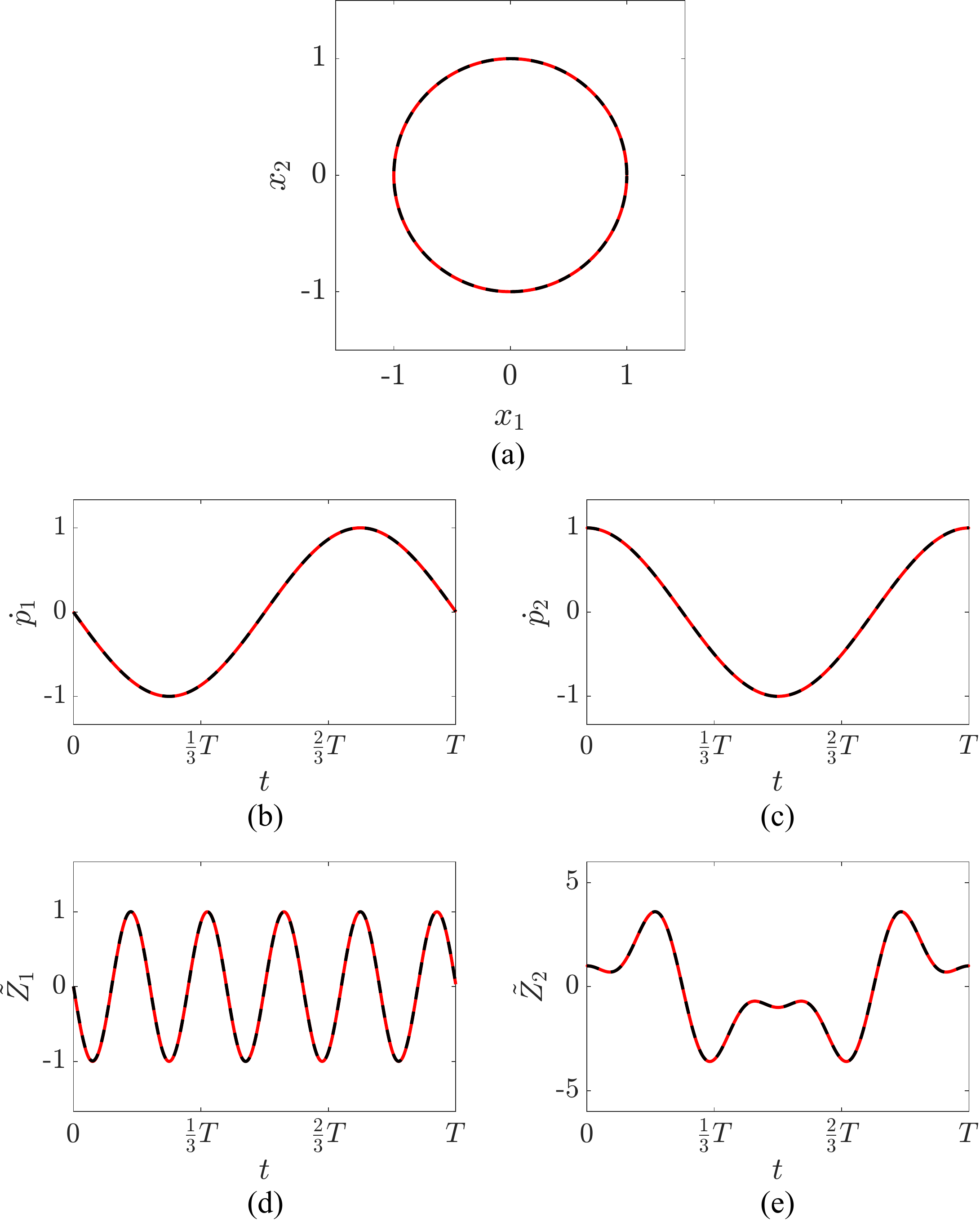}
                    \caption{
                        Periodic trajectory and PSFs of the designed oscillator and the prescribed functional forms.
                        (a)~Periodic trajectories.
                        (b,c)~Velocities on the periodic trajectories.
                        (b)~$x_{1}$ component, (c)~$x_{2}$ component.
                        (d,e)~PSFs.
                        (d)~$x_{1}$ component, (e)~$x_{2}$ component.
                        In each graph, the red line shows the designed functional form and the black dotted line shows the original one, respectively.
                    }
                    \label{fig:namur9}
                \end{figure}

\begin{figure}[t]
                    \centering
                    \includegraphics[width=\hsize]{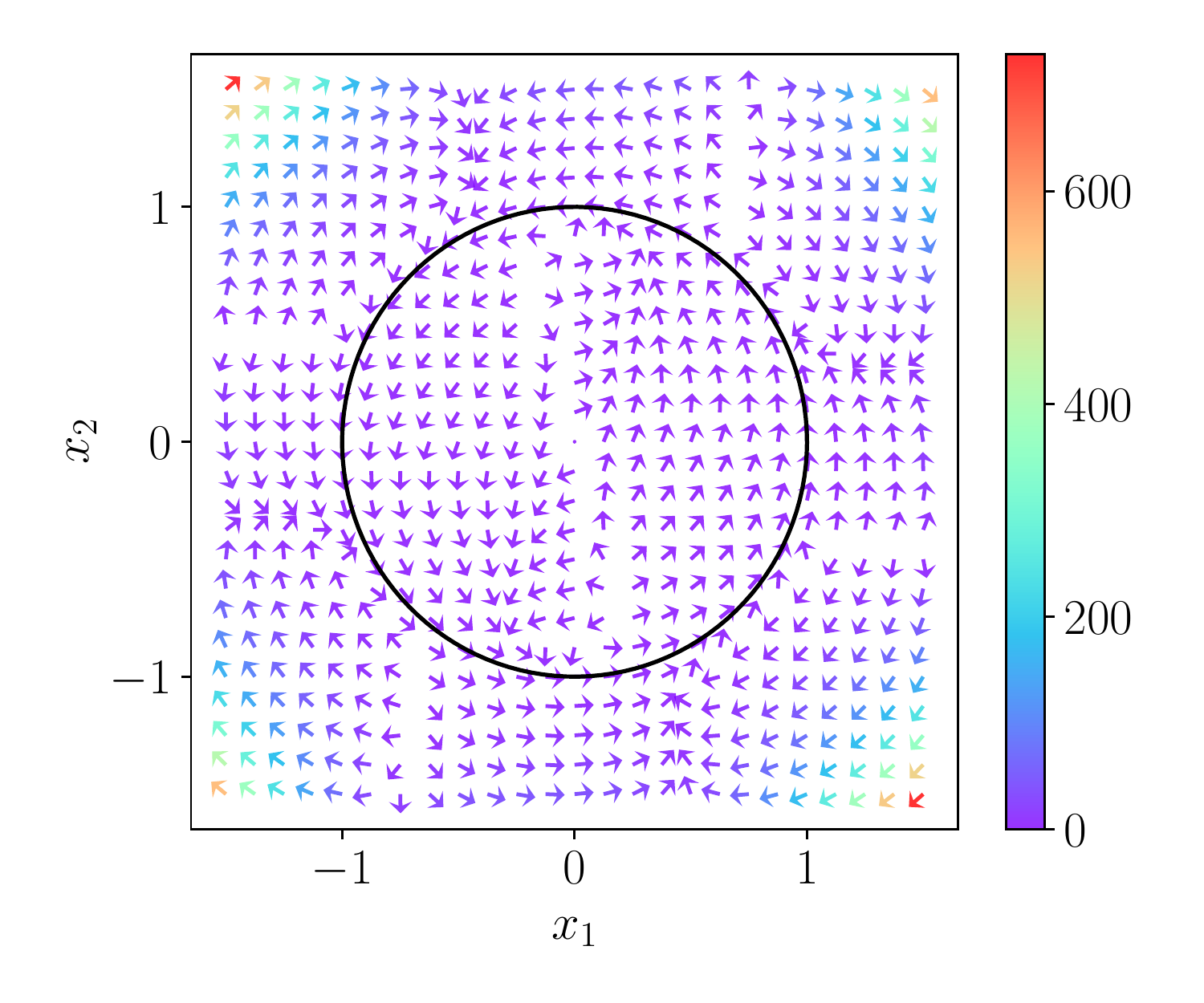}
                    \caption{Vector field and limit cycle of the designed oscillator with an artificial PSF.}
                    \label{fig:namur10}
                \end{figure}

                The limit cycle of the designed oscillator is compared with the given periodic trajectory in Figs.~\ref{fig:namur9}~(a), (b), and (c).
                Both the periodic trajectory and velocity are almost identical to the assumed ones.
                The PSF of the designed oscillator is shown in Figs.~\ref{fig:namur9}~(d) and (e),
                which is also almost identical to the prescribed one.
                The vector field of the designed oscillator is shown in Fig.~\ref{fig:namur10}.

            \subsubsection{Multistable entrainment}

                Here, we demonstrate multistable entrainment of the oscillator designed above. 
                We apply an external periodic input with a frequency $\Omega=5\omega$, which is five times the natural frequency $\omega$. 
                Assuming that the external periodic input has the form of $\bq(t) = \left[ \sin(\Omega t), 0 \right]^{\top}$, we can write the reduced phase equation of the oscillator as
                \begin{align}
                    \ddt{} {\theta(t)} = \omega + \varepsilon \bZ(\theta{(t)}) \cdot \bq(t).
                \end{align}

\begin{figure}[t]
                    \centering
                    \includegraphics[width=0.75\hsize]{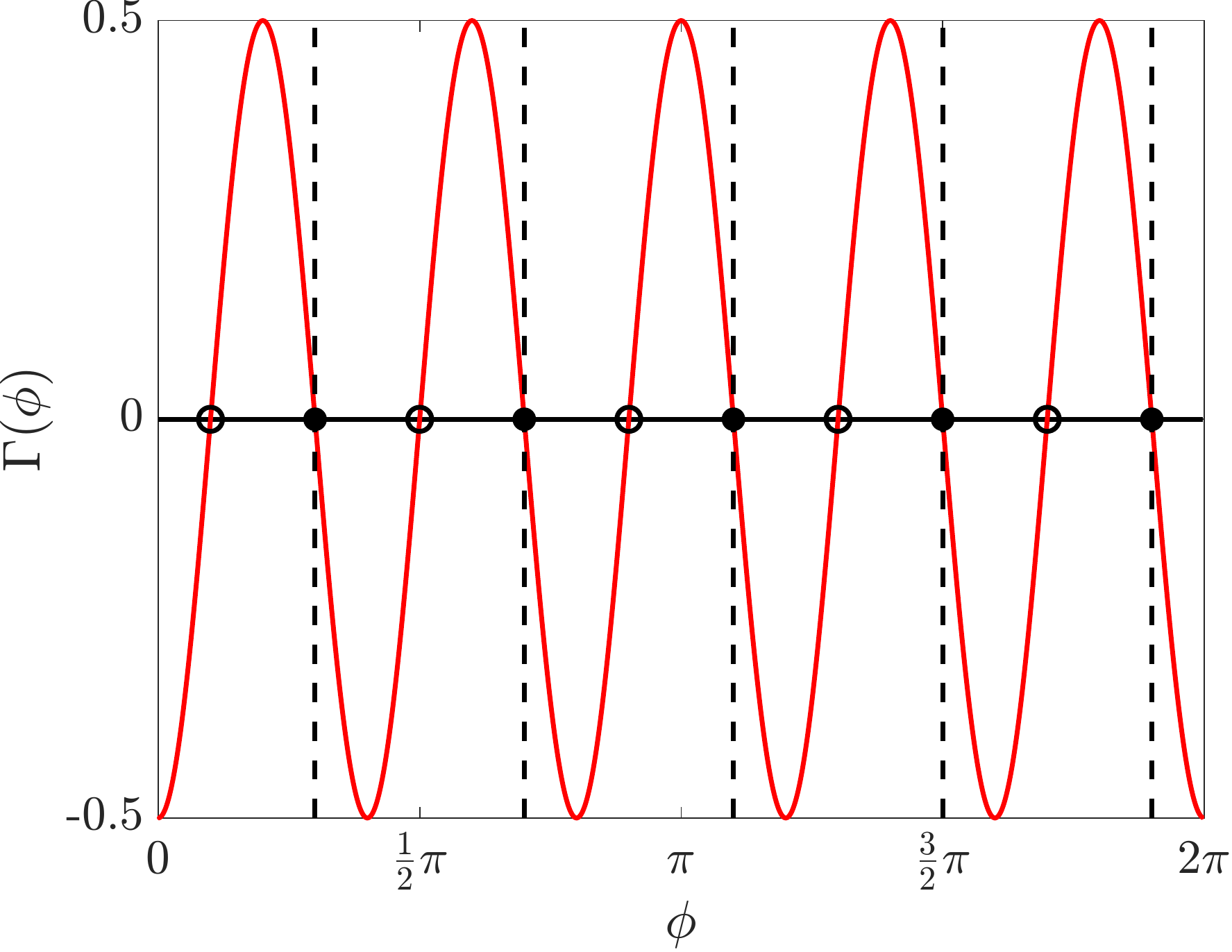}
                    \caption{
                        Phase coupling function $\Gamma(\phi)$ for multistable entrainment.
                        The black solid points show the stable fixed points and the black circle points show the unstable fixed points.
                    }
                    \label{fig:namur11}
                \end{figure}

                Since the ratio of the frequency of the external periodic input to that of the oscillator is $5$, $5:1$ entrainment is expected to occur.
                Generalizing the analysis in Sec.~\ref{sec:theory} and defining the phase of the external input to be $\Omega t/5$, we introduce the phase difference
                \begin{align}
                    \phi{(t)} = \theta{(t)} - \frac{1}{5}\Omega t.
                \end{align}
                The time evolution of $\phi{(t)}$ is then
                \begin{align}
                    \begin{aligned}
                        \ddt{} {\phi(t)} &= \ddt{} {\theta(t)} - \frac{1}{5}\Omega \\
                        &= \omega - \frac{1}{5}\Omega + \varepsilon \bZ \left( \phi{(t)} + \frac{1}{5}\Omega t \right) \cdot \bq(t) \\
                        &= \varepsilon \bZ \left( \phi{(t)} + \frac{1}{5}\Omega t \right) \cdot \bq(t),
                    \end{aligned}
                \end{align}
                and by applying the averaging approximation~\cite{Kuramoto1984,Hoppensteadt1997,Nakao2016}, $\phi{(t)}$ obeys
                \begin{align}
                    \begin{aligned}
                        \ddt{} {\phi(t)} &= \varepsilon\Gamma(\phi{(t)}) \\
                        &= \frac{\varepsilon}{T} \int_{0}^{T} \bZ \left( \phi{(t)} + \frac{1}{5}\Omega t{'} \right) \cdot \bq(t{'}) dt{'} \\
                        &= - \frac{\varepsilon}{2} \cos(5\phi{(t)}).
                    \end{aligned}
                \end{align}
                The phase coupling function $\Gamma(\phi)$ is shown in Fig.~\ref{fig:namur11}.
                There are ten fixed points (five stable and five unstable) satisfying $\dot{\phi} = 0$ within $\phi \in [0,2\pi)$;
                the five stable fixed points are $\phi{^{*}} = 3\pi/10,\; 7\pi/10,\; 11\pi/10,\; 3\pi/2$, and $19\pi/10$.

\begin{figure}[t]
                    \centering
                    \includegraphics[width=\hsize]{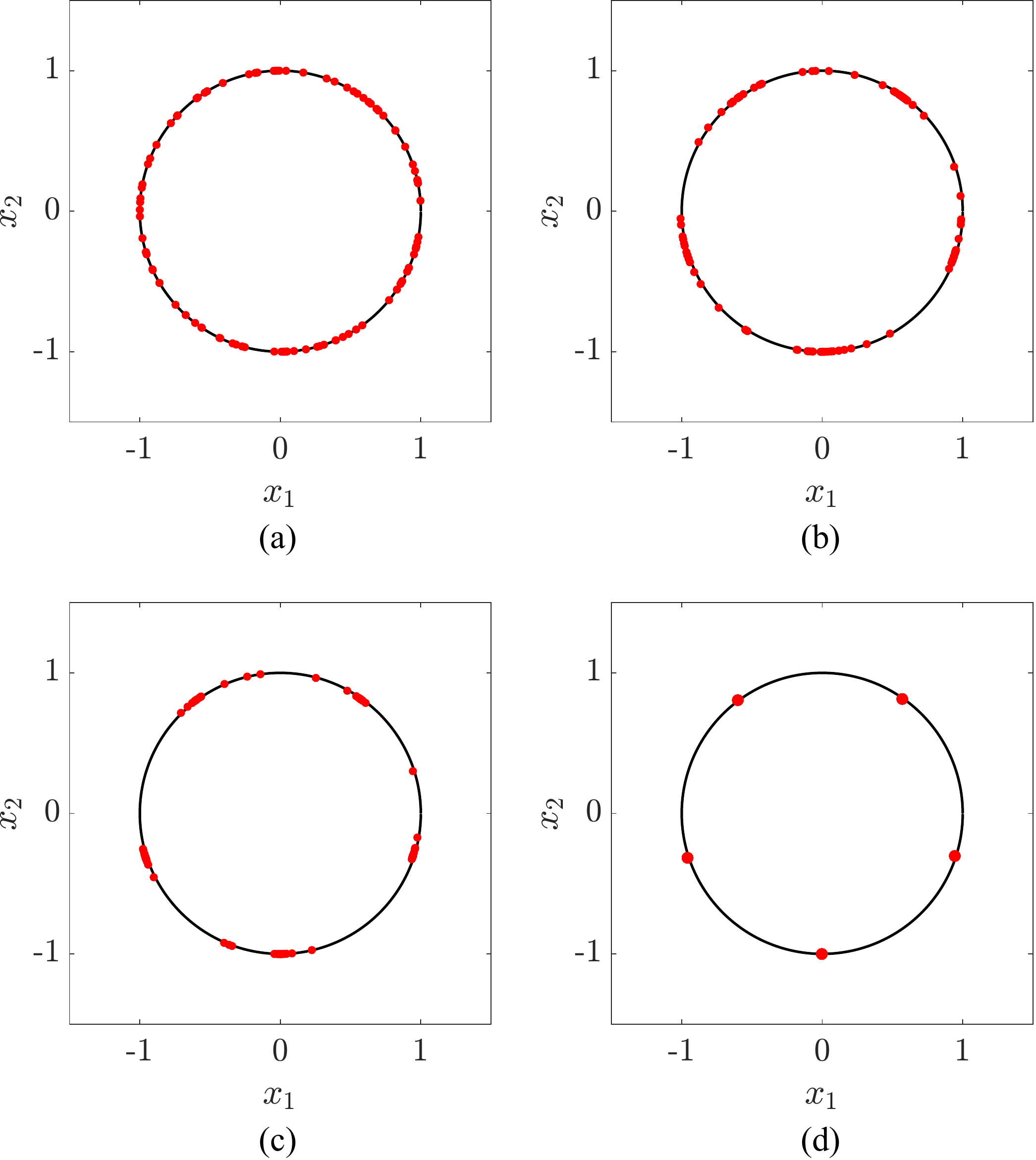}
                    \caption{
                        Multistable entrainment of $N=100$ designed artificial oscillators.
                        {(a)~$t = 0$, (b)~$t = 2\pi$, (c)~$t = 4\pi$, and (d)~$t = 16\pi$.}
                        In (d), we can see the convergence to the five stable fixed points.
                    }
                    \label{fig:namur12}
                \end{figure}

                We performed numerical simulations to confirm that entrainment with multiple phase differences occurs.
                We assumed $\varepsilon = 0.05$ and evolved $100$ independent oscillators from $100$ random initial points on the unit circle.
                The oscillator states at time $t = 0,\; 2\pi,\; 4\pi$, and $16\pi$ are shown in Fig.~\ref{fig:namur12}.
                As time passes, the phases of the oscillators gradually form clusters, and at $t = 16\pi$, the oscillators almost converge to either of the five points.

                We note that synchronization with such many fixed points is rarely observed in natural limit-cycle oscillators. 
                Thus, by using the present method, we can design artificial oscillators with desirable synchronization properties.


    \section{CONCLUDING REMARKS}
    
    \label{sec:conclusion}

        We proposed a method for designing two-dimensional limit-cycle oscillators that possess stable prescribed periodic trajectories and PSFs.
        Using this method, we could design the vector fields that exhibit the periodic trajectories and PSFs of several types of existing or artificial oscillators.
        Furthermore, we were able to design an artificial oscillator with a high-harmonic PSF that rarely exists in the real world, and were able to demonstrate multistable entrainment caused by a high-frequency periodic input.
        The proposed method is simple and generally applicable to design the vector fields of various two-dimensional oscillators. Generalizations of the proposed method to higher-dimensional oscillators will be reported in our future study.


    \acknowledgments
    H. N. thanks financial support from JSPS KAKENHI (Nos. JP22K11919, JP22H00516, JPJSBP120202201) and JST CREST (No. JP-MJCR1913). 
    


\end{document}